# Pulsed THz radiation due to phonon-polariton effect in [110] ZnTe crystal

Chien-Ming Tu,[1,*] Shin-An Ku,[2] Wei-Chen Chu,[2] Chih-Wei Luo,[2] Jeng-Chung Chen,[1] and Cheng-Chung Chi[1]

[1]*Department of Physics, National Tsing Hua University, Hsinchu, Taiwan 30013, R.O.C.*
[2]*Department of Electrophysics, National Chiao Tung University, Hsinchu, Taiwan 30010, R.O.C.*
[*]*d937305@oz.nthu.edu.tw*

**Abstract:** Pulsed terahertz (THz) radiation, generated through optical rectification (OR) by exciting [110] ZnTe crystal with ultrafast optical pulses, typically consists of only a few cycles of electromagnetic field oscillations with a duration about a couple of picoseconds. However, it is possible, under appropriate conditions, to generate a long damped oscillation tail (LDOT) following the main cycles. The LDOT can last tens of picoseconds and its Fourier transform shows a higher and narrower frequency peak than that of the main pulse. We have demonstrated that the generation of the LDOT depends on both the duration of the optical pulse and its central wavelength. Furthermore, we have also performed theoretical calculations based upon the OR effect coupled with the phonon-polariton mode of ZnTe and obtained theoretical THz waveforms in good agreement with our experimental observation.

## 1. Introduction

Terahertz (THz) electromagnetic pulses generated from nonlinear crystals by ultrafast optical pulses have been found to be very useful in numerous fundamental studies and practical applications [1]. The THz pulse radiation, most commonly generated through optical rectification (OR) by exciting [110] ZnTe crystal with ultrafast optical pulses, typically consists of only a few cycles of electromagnetic field oscillations ("few-cycle electromagnetic oscillations") with a duration about a couple of picoseconds [2]. To our knowledge, we have found only one instance where it was reported that THz waveform consists of not only a few-cycle pulse but also a long damped oscillation tail (LDOT) [3]. Although the evolution of THz waveforms with the central wavelength $\lambda_0$ of the optical pulse was demonstrated by Ahn *et al.*, the detailed mechanism for generating the LDOT has not been fully addressed.

In polar materials such as LiNbO$_3$ and ZnTe, it is well-known that the interaction of a THz electromagnetic wave with a TO phonon of comparable frequency leads to a mixed mode, the so-called phonon-polariton. Recently, Wahlstrand and Merlin unified difference-frequency generation (or OR) and impulsive stimulated Raman scattering for describing coherent phonon-polariton generation by ultrafast optical pulse inside ZnTe, GaP and LiTaO$_3$ crystals [4]. In the pump-probe experiments, the bandwidth of a single tightly-focused optical pulse can cover a substantial portion of the phonon-polariton frequency spectrum, and the frequency-matching and the wave-vector-matching conditions are possible for strong coherent phonon-polariton generation. However, to our knowledge, no detailed study of free space THz radiation, generated by ultrafast optical pulse illuminating [110] ZnTe and taking into account phonon-polariton effect, has been reported

Here, we report such a study of phonon-polariton effect on THz radiation generated by ultrafast optical pulse in [110] ZnTe. We have demonstrated that phase-matched phonon-polariton can be generated efficiently and greatly influence the THz waveform in free space. In the past, the duration of optical pulse effecting on the THz generation process has been overlooked. The dependence of THz waveform on optical pulse duration gives an insight to the THz generation process. The influences of optical wavelength on THz generation are also clearly observed. In order to understand more quantitatively the generation of LDOT, we have also performed the theoretical calculations based on a nonlinear wave conversion model with both the phonon-polariton dispersion and the second-order nonlinear susceptibility of TO-phonon correction, to compare with the experimental measurements.

## 2. Experiments and results

The laser used was a mode-locked Ti: sapphire oscillator (Tsunami, Spectra Physics) producing 10-nJ optical pulses of central wavelength $\lambda_0$ =750 nm, pulse duration $\Delta t_p$ ~150 fs at a repetition rate of 76 MHz (the cavity length was modified to change the standard repetition rate of 82 MHz to 76 MHz for other purposes). The laser beam was split into a stronger pump beam for THz generation and a weak probe beam for standard electro-optic sampling (EOS). The THz radiation generated by illuminating 2-mm thick [110] ZnTe with the pump beam was collimated and focused by two off-axis parabolic mirrors, and another [110] ZnTe slab of 1-mm thickness was used for THz detection by using EOS. The experimental setup was exposed in ambient air.

Figure 1 shows the measured THz waveform and the corresponding Fast Fourier Transform (FFT) spectrum. In Fig. 1(a), at a laser pulse duration of a full-width-at-half-maximum $\Delta t_p$ ~150 fs and central wavelength $\lambda_0$ =750 nm, the THz waveform is clearly not a typical few-cycle THz pulse and instead is composed of the main THz pulse and a LDOT. As shown in the inset of Fig. 1(a), there are two broad peaks in the corresponding FFT spectrum: the lower-frequency peak centers at $f_L$ ~0.66 THz and the high frequency peak centers at 2.70 THz, corresponding to the main THz pulse and the LDOT, respectively. Coherent length $L_{coh}(\omega)$, defined by $L_{coh}(\omega) = \pi \left( \omega \left| c/V - c/v_{ph}^{P} \right| \right)^{-1}$ ($c$ is the light speed in vacuum, $\omega$ is the angular frequency in THz range, $V$ is the group velocity of the optical pulse, and $v_{ph}^{P}$ is the phase velocity of the phonon-polariton), is also shown in the inset of Fig. 1(a) [5]. The phase-matched frequency $f_{PM}^{750nm}$, determined by $L_{coh}(\omega) \to \infty$, for $\lambda_0$ =750 nm is 2.70 THz, which is exactly the center frequency of the LDOT. Between the two peaks in the spectrum, there is a smaller and broader bump. It is well-known that for the non-phase matched frequencies, the conversion from the optical pulse reaches a local maximum when the crystal length equals the odd integral multiple of the coherence length. In our case, the used crystal thickness is 2 mm, which is approximately triple of the coherent length (about 0.7 mm) for frequencies about 1.5 THz. Therefore, this bump is the local maximum value of non-phase-matched frequencies for this thickness of ZnTe. To further verify the relationship between the phase-matched frequency and the LDOT, we deliberately lengthened the optical pulse width and changed the laser wavelength. In Fig. 1(b), the measured THz waveform, produced by an optical pulse of $\Delta t_p$ ~220 fs, $\lambda_0$ =744 nm, apparently shows only a few-cycle THz pulse in the time domain, but its FFT spectrum, shown in the inset of Fig. 1(b), shows a main peak at 0.50 THz and a smaller and broader bump at about 1.30 THz. However, there is no clear peak at $f_{PM}^{744nm}$ = 2.80 THz but only a residual component in the FFT spectrum.

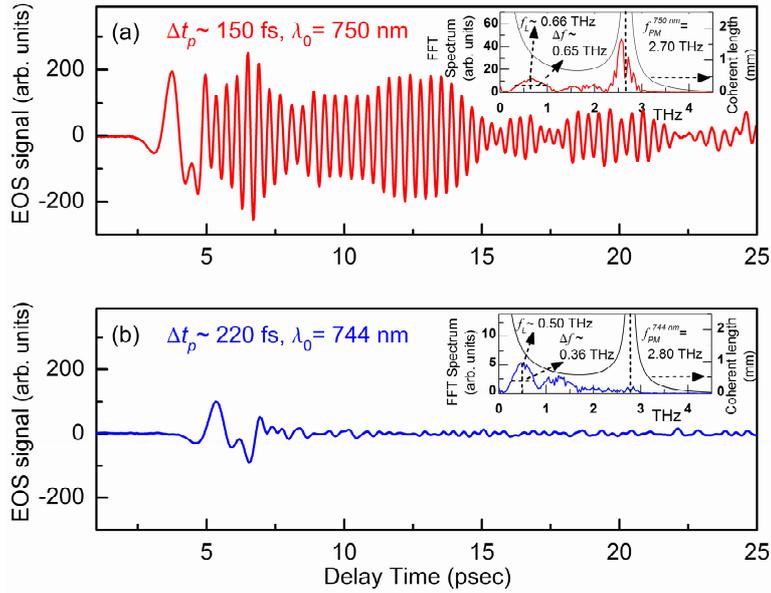

Fig. 1. (a) For optical pulse of $\Delta t_p$ ~150 fs, the measured THz waveform consists of a main THz pulse and a long damped oscillation tail (LDOT). The peak of the phase-matched frequency $f_{PM}^{750nm}$ =2.70 THz (for $\lambda_0$ =750 nm) corresponds to the LDOT. The coherent length of $\lambda_0$ =750 nm is also shown in the inset of (a). (b) For $\Delta t_p$ ~220 fs and $\lambda_0$ =744 nm, only a pure THz pulse was measured in time domain. The LDOT almost disappears and the amplitude of the phase-matched frequency ($f_{PM}^{744nm}$ =2.80 THz for $\lambda_0$ =744 nm) also reduces in the FFT spectrum.

The most common center wavelength $\lambda_0$ of a mode-locked Ti: sapphire laser is around 800 nm, and it has been commonly observed that THz radiation generated from [110] ZnTe by an optical pulse of this wavelength is of the few-cycle pulse type. In order to explore the possibility of observing LDOT with an 800-nm laser pulse, we switched to another two different Ti: sapphire laser systems, both of which produced optical pulses of $\lambda_0 = 800$ nm but with different pulse durations. First, a mode-locked Ti: sapphire oscillator (FEMTOLASERS: Model XL 300), which produced 300-nJ optical pulses of $\Delta t_p \sim 70$ fs at a repetition rate of 5 MHz, was employed and the experiment was conducted in a chamber filled with nitrogen gas to avoid THz-waveform distortion due to water vapor absorption. The measured THz waveform and the FFT spectrum are shown in Fig. 2(a). It can be seen that the measured THz waveform contains not only a few-cycle THz pulse but also a LDOT of several picoseconds. In the inset of Fig. 2(a), there are two broad peaks overlapping in the FFT spectrum, and the 1.90-THz peak meets the phase matching condition of $\lambda_0 = 800$ nm. Because of the short optical pulse used, the spectrum center $f_L$ of the main THz pulse is higher than those of the cases shown in Fig. 1. For the second system, we used a regenerated amplifier (Hurricane, Spectra Physics), which produced optical pulses of $\Delta t_p \sim 150$ fs at a repetition rate of 1 kHz to perform the same experiment. Fig. 2(b) shows the THz wave generated by the regenerated amplifier and the waveform is the typical few-cycle THz pulse. This experiment was performed in ambient air, and some irregular oscillations due to vapor absorption are apparent behind this THz pulse. Because of the broader optical pulses used, the whole spectrum congregates in the low-frequency region, and the frequency component around $f_{PM}^{800nm} = 1.90$ THz almost disappears.

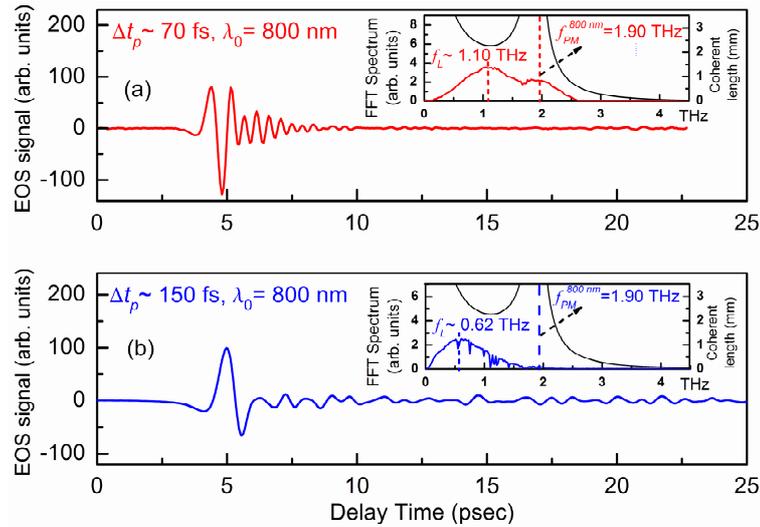

Fig. 2. (a) This experiment was performed in a chamber filled with nitrogen gas with the FEMTOLASERS (Model XL 300) laser system producing optical pulses of $\Delta t_p \sim 70$ fs, $\lambda_0 = 800$ nm. The measured THz waveform is also composed of a few-cycle THz pulse and a LDOT. Two broad peaks overlap in the spectrum and the 1.9-THz peak satisfies the phase-matching condition. (b) This experiment was performed in ambient air with a regenerative amplifier Ti:sapphire laser producing pulses of $\Delta t_p \sim 150$ fs, $\lambda_0 = 800$ nm. The few-cycle THz pulse with some irregular oscillations behind it was observed.

## 3. Theoretical calculations and discussions

Phonon-polariton is a hybrid mode of THz wave and TO phonon of polar material. The polarization due to infrared-active TO phonon also contributes to the polarization wave of a THz wave traveling in polar media [6]. In the following, we refer to phonon-polariton as polariton for abbreviation. For ZnTe, the dielectric function pertaining to the polariton dispersion is given by

$$\frac{c^2 k^2(\omega)}{\omega^2} = \varepsilon(\omega), \varepsilon(\omega) = \varepsilon(\infty) + \frac{[\varepsilon(0) - \varepsilon(\infty)]\omega_{TO}^2}{\omega_{TO}^2 - \omega^2 + i\gamma\omega}. \quad (1)$$

where $k(\omega)$ is the wave-vector of polariton, $\varepsilon(0)$ and $\varepsilon(\infty)$ are the static- and the high-frequency dielectric constants, respectively, $\omega_{TO}$ is the frequency of the TO phonon [7], and $\gamma$ is the damping rate of the TO phonon. Here, we set $\gamma/2\pi = 0.025$ THz [8]. For ZnTe, the refractive index in THz range can be expressed as $n^{THz} = \sqrt{\varepsilon(\omega)}$ and the phase velocity $v_{ph}^P$ of the polariton is $v_{ph}^P = \omega/k = c/\sqrt{\varepsilon(\omega)} = c/n^{THz}$.

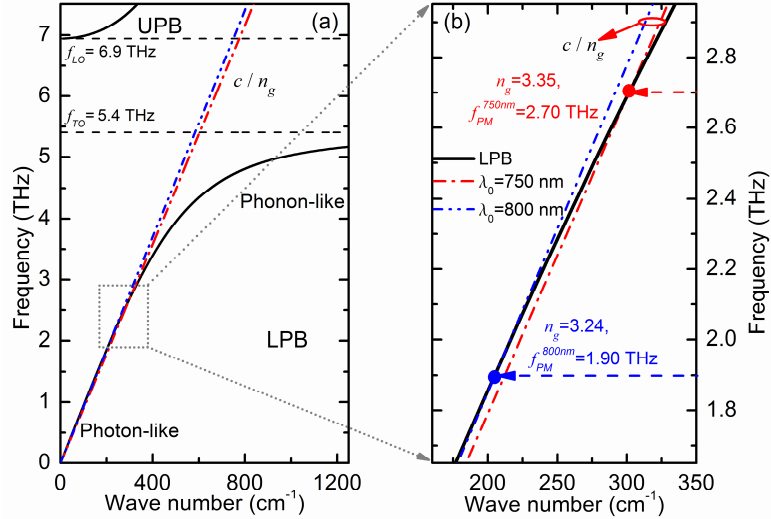

Fig. 3. (a) Phonon-polariton dispersion of ZnTe. (b) The detail of the LPB in middle region. The light lines representing optical pulses of $\lambda_0 = 750$ nm and $\lambda_0 = 800$ nm intersect the LPB, and the intersections represent the phase-matched frequencies of $f_{PM}^{750nm} = 2.70$ THz and $f_{PM}^{800nm} = 1.90$ THz, respectively.

Figure 3 demonstrates the polariton dispersion of ZnTe without $\gamma$. The slopes of the straight lines represent the group velocities of the optical pulses of $\lambda_0 = 750$ nm and $\lambda_0 = 800$ nm inside ZnTe. These two lines intersect the lower polariton branch (LPB) at frequencies of 2.70 THz and 1.90 THz, respectively, which define the phase-matching conditions, as shown in Fig. 3 (b). In the LPB, as the frequency of the polariton approaches that of a TO phonon, the polariton carries more characteristics of a TO phonon and its propagation speed is also affected dramatically.

To compare with the experimental results, we calculated both waveforms and spectra of THz radiation in free space. Considering planar wave propagation in the $z$-direction and the pump non-depletion situation, the equation of THz wave generation including the effect of polariton by single optical beam can be expressed as [4,9]

$$\frac{\partial^2}{\partial z^2}\tilde{E}^{THz}(z,\omega)+\frac{\omega^2}{c^2}\varepsilon(\omega)\tilde{E}^{THz}(z,\omega)=(-1)\frac{4\pi\omega^2}{c^2}\chi^{(2)}(\omega)\tilde{F}(\omega)\exp\left(\frac{-i\omega z}{V}\right). \quad (2)$$

where $\tilde{E}^{THz}(z,\omega)$ is the amplitude of the THz E-field in the frequency domain. $V=c/n_g$, where $n_g$ is the group index of the optical pulse in ZnTe. In the near-infrared range, the group index $n_g$ of ZnTe is obtained from Ref. 10. $\chi^{(2)}(\omega)$ is the second-order nonlinear susceptibility of ZnTe in the THz frequency range and it also shows strong dispersion in the vicinity of $\omega_{TO}$ [4,11]. The function $\tilde{F}(\omega)$ in Eq. (2) is the frequency spectrum of Fourier transform-limited optical pulse, and the intensity of the optical pulse is normalized to give $\tilde{F}(\omega)=\exp(-\omega^2\tau_G^2/4)$, where $\tau_G$ is the Gaussian width and is related to the full-width-at-half-maximum of the optical pulse duration $\Delta t_p$ by $\Delta t_p = 2\sqrt{\ln 2}\;\tau_G$. For a THz-emitter ZnTe of thickness $L$, we used the boundary conditions across both incident and exit interfaces to obtain the THz waveform in the free space [12], i.e. $z>L$

$$E^{THz}(z>L,t)=\frac{1}{2\pi}\int_{-\infty}^{\infty}d\omega\,\tilde{E}^{THz}(z=L,\omega)e^{i\omega[t-(z-L)/c]}$$

$$\tilde{E}^{THz}(z=L,\omega)=\frac{4\pi\chi^{(2)}(\omega)\tilde{F}(\omega)}{\varepsilon(\omega)-n_g^2}\times$$

$$\frac{2\sqrt{\varepsilon(\omega)}(1+n_g)+\left(\sqrt{\varepsilon(\omega)}-1\right)\left(\sqrt{\varepsilon(\omega)}-n_g\right)e^{-i\frac{\omega L}{c}\left[n_g+\sqrt{\varepsilon(\omega)}\right]}-\left(\sqrt{\varepsilon(\omega)}+1\right)\left(\sqrt{\varepsilon(\omega)}+n_g\right)e^{-i\frac{\omega L}{c}\left[n_g-\sqrt{\varepsilon(\omega)}\right]}}{\left(\sqrt{\varepsilon(\omega)}+1\right)^2 e^{i\frac{\omega L}{c}\sqrt{\varepsilon(\omega)}}-\left(\sqrt{\varepsilon(\omega)}-1\right)^2 e^{-i\frac{\omega L}{c}\sqrt{\varepsilon(\omega)}}}.$$

(3)

Based on Eq. (3), we calculated THz waveforms and spectra under various experimental conditions. Using $L=2$ mm, Fig. 4(a) shows calculated THz waveforms excited by optical pulses of $\lambda_0=750$ nm with pulse durations of $\Delta t_p=150$, 220 and 450 fs. The calculated THz waveforms clearly consist of a main THz pulse and a LDOT, similar to our experimental observation, and the LDOT diminishes as optical pulse duration increases. The double-reflected THz waves inside a ZnTe emitter of 2-mm are also shown. The corresponding spectra are shown in Fig. 4(b), and the higher frequency peaks for the three optical pulses, centering at $f_{PM}^{750nm}=2.70$ THz, represent the LDOTs in the spectrum. The amplitude of this peak reduces as the optical pulse duration $\Delta t_p$ is broadened from 150 to 450 fs. Between the phase-matched peak and the low frequency peak corresponding the main pulse, the calculated spectra also shows a broad bump, consistent with our experimental observations.

As shown in Fig. 3(a) and 3(b), the LDOT of 2.70-THz matches the intersection point of the LPB very well. We calculated both the phase-velocity $v_{ph}^P$ (=0.30 $c$) and the group velocity $v_{gr}^P$ (=0.26 $c$) of the polariton at this intersection, and the estimated time-delay from the optical pulse for a 2.70-THz polariton wave-packet traveling through 2-mm ZnTe is about 3.4 psec. Indeed, this time-delay is comparable with the duration of the calculated LDOT, as shown in Fig. 4(a). It is clear that the polariton generation becomes very efficient when the phase-matching condition is satisfied in ZnTe, but the generated polariton wave packet tails behind the optical pulse front to produce the observed LDOT.

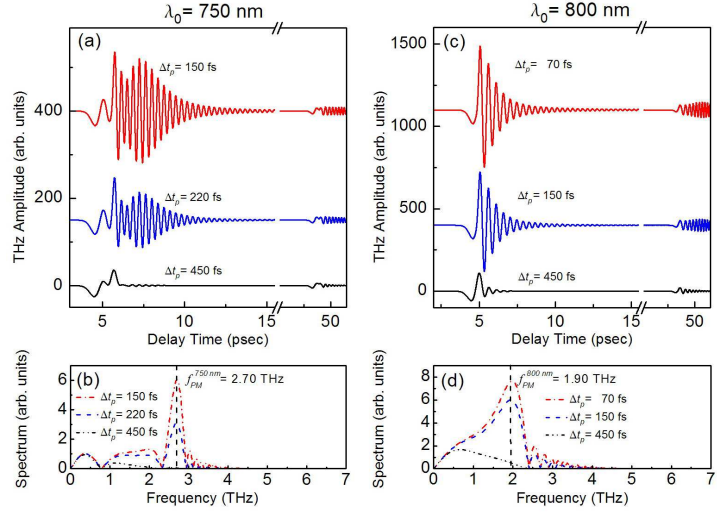

Fig. 4. (a) Calculated THz waveforms excited by optical pulses of $\lambda_0$ =750 nm with $\Delta t_p$ =150, 220 and 450 fs. The corresponding spectra are also shown in (b). (c) Calculated THz waveforms excited by optical pulses of typical $\lambda_0$ =800 nm with $\Delta t_p$ =70, 150 and 450 fs, and the corresponding spectra are shown in (d). With the broadening of optical pulse duration, the amplitude of the phase-matched peak reduces; the calculated LDOT shrinks in the time domain and its amplitude also reduces.

The most obvious discrepancy between the observed and the calculated LDOT is that the former persists longer and shows some envelope modulation as shown in Fig. 1(a). The main reason for this discrepancy is generally believed due to some water vapor absorptions, which shows up in the experimental spectrum [13], as the absorption dips in the vicinity of the frequency of 2.7 THz (see inset of Fig. 1(a)). To verify that the discrepancy is indeed due to water vapor absorption, we also calculate the THz waveform and spectrum by taking into account of the several water vapor absorption lines around 2.70 THz. As shown in Fig. 5, the peak centering at 2.70 THz shows some absorption dips, and the corresponding LDOT becomes longer and indeed shows some envelope modulation. These results agree well with our experimental observations.

In our experimental observations, the broadening of the optical pulse also affects the main THz pulse. The bandwidth $\Delta f$ of the broad peak of low-frequency, which represents the main THz pulse, becomes narrower from 0.65 THz to 0.36 THz when the duration of optical pulse is broadened from 150 fs to 220 fs (see the insets of Fig. 1(a) and 1(b)). The corresponding spectrum center peak frequency $f_L$ of the main THz pulse also decreases from 0.66 THz to 0.50 THz.

During actual operation of the laser, it is difficult to adjust the central wavelength and the duration of the optical pulse independently. In Fig. 1, during adjustments to the optical pulse duration, the central wavelength shifts slightly by about 6 nm (750-744 = 6 nm) and the phase-matched frequency of $\lambda_0$ =744 nm shifts to 2.80 THz. This small difference in the phase-matched frequency ($\Delta f_{PM}$ =2.80-2.70=0.1 THz) has little impact on the characteristics of the excited polariton. Thus, the reduced amplitude of the LDOT is mostly due to the increase of the optical pulse duration. The reason for this dependence is now clear. The magnitude of the Fourier component near the phase-matched frequency becomes diminishingly small as the optical pulse width increases.

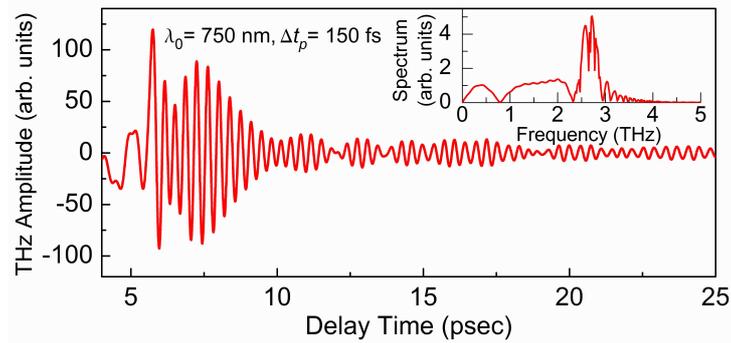

Fig. 5. Calculated THz waveform and spectrum in consideration of water vapor absorption lines around 2.7 THz for $\Delta t_p$ =150 fs and $\lambda_0$ =750 nm. Some envelope modulations appear on the LDOT apparently.

For comparison, we have also calculated the THz waveforms and the corresponding spectra for optical pulses of $\lambda_0$ =800 nm with different pulse durations. As shown in Fig. 4(c) and 4(d), LDOTs indeed appear in the calculated THz waveforms and the broad peaks at $f_{PM}^{800nm}$ =1.90 THz appear in the spectra. The broadening of the optical pulse also strongly affects the calculated waveforms and spectra.

Compared to the case of $\lambda_0$ =750 nm, the phase-mated intersection of $\lambda_0$ =800 nm in the polariton dispersion (see Fig. 3) is closer to the photon-like region, and the difference between the corresponding phase-velocity and the group-velocity of the polariton (for 1.90 THz, $v_{ph}^P$ =0.31 $c$ and $v_{gr}^P$ =0.30 $c$) is smaller than that in the higher frequency region of the LPB (for 2.70 THz, $v_{ph}^P$ =0.30 $c$ and $v_{gr}^P$ =0.26 $c$). Thus, the duration of the LDOT of 1.90-THz is shorter than that of 2.70-THz. A similar phenomenon has also been reported in the pump-probe measurement of ZnSe [14]. Our works has demonstrated the detailed conditions for generating tunable-frequency THz radiation by changing the wavelength of optical pulses. Frequency-tuning of THz radiation utilizing phonon-polariton dispersion in LiNbO$_3$ also has been reported, and that technique, tilted optical pulse front to achieve non-collinear phase-matching in LiNbO$_3$, has been shown great impact on high power THz generation [15].

In theoretical calculations, we consider only the absorption due to the TO phonon by using the imaginary part of the refractive index. There are some differences between the experimental results and the numerical calculations, especially in the case of broad-optical-pulse excitation. For ZnTe, there are strongly high-order absorption processes in the high frequency range [8,16] and the chirp of the optical pulse is not included in our calculations. These effects may be responsible for the remaining discrepancy between the measured and calculated LDOT. Furthermore, the THz-waveform distortion due to EOS is not included in all calculations, and far-field diffraction for a finite ZnTe slab is not contained [17].

## 4. Conclusions

In conclusion, we have demonstrated the phonon-polariton effect on THz generation by optical pulses in [110] ZnTe crystal. Both experimental measurements and theoretical calculations show that the THz waveforms are composed of a main THz pulse and a long-damped oscillation tail. The long-damped-oscillation tail is the result of the phonon-polariton effect inside [110] ZnTe through phase-matching. The optical-pulse-duration dependence of the long damped oscillation tail has been confirmed both experimentally and theoretically. As the duration of the optical pulse becomes broad enough, the amplitude of the long damped oscillation tail becomes smaller and eventually only a few-cycle THz pulse remains in time domain. Our work provides a clearer and quantitative picture of THz generation in [110] ZnTe, and of their dependences on laser pulse central frequency and pulse duration.

**Acknowledgements**

This work was supported by National Science Council, Taiwan, R.O.C. Grant No. 99-2120-M-007-002.